\documentclass[twocolumn,english,aps,prl,superscriptaddress,10pt]{revtex4-1}
\usepackage[T1]{fontenc}
\usepackage[latin9]{inputenc}
\usepackage[a4paper]{geometry}
\usepackage{color}
\usepackage{babel}
\usepackage{amsmath}
\usepackage{amssymb}
\usepackage{graphicx}
\usepackage{esint}
\usepackage[unicode=true,pdfusetitle,
 bookmarks=true,bookmarksnumbered=false,bookmarksopen=false,
 breaklinks=false,pdfborder={0 0 1},backref=false,colorlinks=false]
 {hyperref}
\usepackage{breakurl}

\makeatletter
\@ifundefined{textcolor}{}
{%
 \definecolor{BLACK}{gray}{0}
 \definecolor{WHITE}{gray}{1}
 \definecolor{RED}{rgb}{1,0,0}
 \definecolor{GREEN}{rgb}{0,1,0}
 \definecolor{BLUE}{rgb}{0,0,1}
 \definecolor{CYAN}{cmyk}{1,0,0,0}
 \definecolor{MAGENTA}{cmyk}{0,1,0,0}
 \definecolor{YELLOW}{cmyk}{0,0,1,0}
}

\usepackage{babel}
\usepackage{babel}
\usepackage{babel}
\usepackage{babel}
\usepackage{babel}

\makeatother

\begin{document}

\title{Space-time vortex driven crossover and vortex turbulence phase transition in one-dimensional
driven open condensates}

\author{Liang He}
\affiliation{Institute for Theoretical Physics, Technical University Dresden, D-01062 Dresden, Germany}
\affiliation{Institute for Theoretical Physics, University of Cologne, D-50937 Cologne, Germany}

\author{Lukas M. Sieberer}
\affiliation{Department of Condensed Matter Physics, Weizmann Institute of Science, Rehovot 7610001, Israel}
\affiliation{Department of Physics, University of California, Berkeley, California 94720, USA}

\author{Sebastian Diehl}
\affiliation{Institute for Theoretical Physics, Technical University Dresden, D-01062 Dresden, Germany}
\affiliation{Institute for Theoretical Physics, University of Cologne, D-50937 Cologne, Germany}

\begin{abstract}
We find a first order transition driven by the strength of non-equilibrium conditions of one-dimensional driven open condensates. Associated with this transition is a new stable non-equilibrium phase, space-time vortex turbulence, whose vortex density and quasiparticle distribution show strongly non-thermal behavior. Below the transition, we identify a new time scale associated with noise activated unbound space-time vortices, beyond which the temporal coherence function changes from a Kardar-Parisi-Zhang type subexponential to a disordered exponential decay. Experimental realization of the non-equilibrium vortex turbulent phase is facilitated in driven open condensates with a large diffusion rate. 
\end{abstract}

\maketitle

The last decade has witnessed fast experimental development in realizing driven
open quantum systems with many degrees of freedom. Examples include
exciton-polaritons in semiconductor
heterostructures~\cite{Polarition_exp_1,Polarition_exp_2,Polarition_exp_3},
ultracold
atoms~\cite{neq_cold_atom_exp_1,neq_cold_atom_exp_2,neq_cold_atom_exp_3},
trapped ions~\cite{neq_trapped_ion_1,neq_trapped_ion_2}, and microcavity
arrays~\cite{neq_micro_cavities_1,neq_micro_cavities_2}. The common
characteristic is explicit breaking of detailed balance on a microscopic level
by the presence of both coherent and driven-dissipative dynamics, placing these
systems far from thermal equilibrium. This makes them promising laboratories for
studying non-equilibrium statistical mechanics, according to which one expects
non-equilibrium features to persist to the macroscopic level of observation.

A case in point are exciton-polariton systems, which can be engineered in one-
and two-dimensional
geometries~\cite{Polarition_exp_1d,Polarition_exp_1,Polarition_exp_2,Polarition_exp_3}. Due
to effectively incoherent pumping, these systems possess a phase rotation symmetry
and can thus show Bose condensation phenomena. While their dynamics is described
microscopically in terms of a stochastic complex Ginzburg-Landau equation (SCGLE)~\cite{Carusotto2013}, 
recently it was noticed that at low
frequencies it maps to the Kardar-Parisi-Zhang (KPZ)
equation~\cite{Altman_2d_driven_SF_2015, Wouters_1d_spatial}. Traditionally, the latter equation
describes, e.g., the roughening of surfaces, with the dynamical surface height
being unconstrained~\cite{KPZ_equation}. In contrast, the dynamical variable in
the context of driven open quantum systems is the condensate phase, which is
compact. The compact KPZ (cKPZ) equation gives rise to a novel scenario for
non-equilibrium statistical mechanics realized in concrete experimental
platforms, and raises the fundamental question of the physical consequences of
compactness.

\begin{figure}
\includegraphics[width=3.4in]{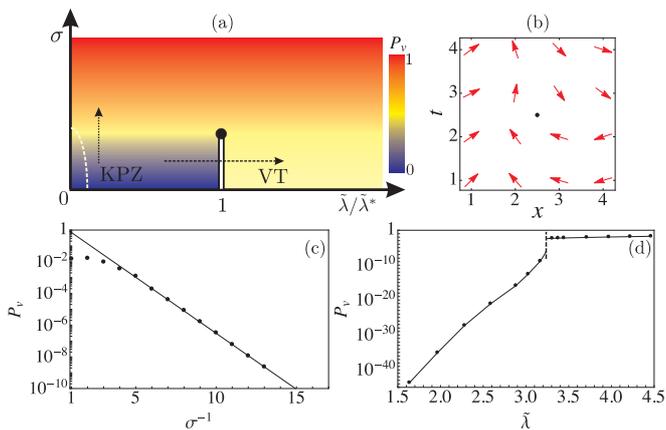}
\caption{(Color online) (a) Schematic phase diagram of a generic 1D driven open
  condensate (DOC) with noise level $\sigma$ and rescaled non-equilibrium
  strength parameter $\tilde{\lambda}/\tilde{\lambda}^{*}$. The color code
  stands for space-time vortex density $P_v$. A first order phase transition
  at low noise level separates a regime dominated by KPZ physics from a vortex
  turbulent (VT) regime.  At stronger noise, the first order
    transition line (double line) terminates at a second order critical point
    (filled black circle, cf.~Figs.~\ref{Fig. fig_3}(c) and \ref{Fig. fig_3}(d)
    for quantitative results). Current exciton-polariton condensate experiments
  are located within the white dashed arc. (b) A typical phase configuration on
  the space-time plane corresponding to a phase-slip event between $t=2$ and
  $3$. The space-time vortex core is marked with a black dot. (c) Noise level
  dependence of the space-time vortex density $P_v$ at small non-equilibrium
  strength $\tilde{\lambda}<\tilde{\lambda}^{*}$ (vertical dashed arrow in
  (a)). At low noise level, $P_v\propto e^{-A/\sigma}$ ($A$ a non-universal
  positive constant) is suppressed exponentially, reflected by the linear fit
  for $\sigma^{-1}=6,...,13$. Values of other parameters used in simulations are
  $K_{d}=r_{d}=u_{d}=1,r_{c}=u_{c}=0.1$, and $K_{c}=3$. (d) Non-equilibrium
  strength dependence of $P_v$ at low noise level with $\sigma=10^{-2}$,
  demonstrating the first order transition upon increasing the non-equilibrium
  strength $\tilde{\lambda}$ (horizontal dashed arrow in (a)). Values of other
  parameters used in simulations are $K_{d}=r_{d}=u_{d}=1$, $K_{c}=0.1$,
  and $\tilde{\lambda}$ is tuned by changing $r_{c}=u_{c}$ from
    $1.0$ to $3.0$. See text for more details.}
\label{Fig. Phase_diagram_and_key_properties} 
\end{figure}

\begin{figure}
\includegraphics[width=3.3in]{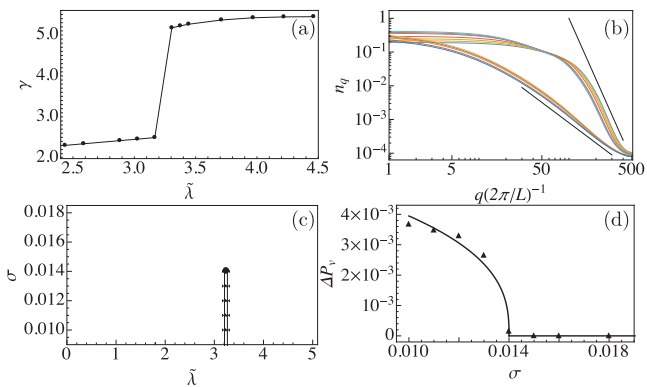}
\caption{ (Color online) (a) Scaling behavior of the momentum
    distribution $n_{q}\propto q^{-\gamma}$ at large $q$ and for different
    $\tilde{\lambda}$ at weak noise ($\sigma=10^{-2})$, showing a transition
    from $\gamma \simeq 2$ to $\gamma \simeq 5$.  (b) Corresponding momentum
    distribution from which the value for different $\gamma$ in (a) is
    obtained. The upper and lower black lines correspond to $\gamma = 5$ and
    $\gamma = 2$, respectively. From right to left, the curves in between the
    two black lines correspond to $r_{c}=u_{c}$ decreasing from $3.0$ to $1.5$.
    (c) $\tilde\lambda^*$ at different noise level with corresponding error bar
    ($\tilde\lambda^*=3.23\pm 0.07$ for all $\sigma$'s shown in the plot). The
    first order transition line (double line) terminates in a second order
    transition point at a $\sigma^*\simeq0.014$ (black filled circle).  (d)
    Space-time vortex density jump $\Delta P_v$ at
    $\tilde\lambda=\tilde\lambda^*=3.23$ at different $\sigma$. The part of the
    black curve at $\sigma\leq \sigma^*$ is a power law
    ($\Delta P_v\propto (\sigma-\sigma^*)^{1-\kappa}$) fit, which yields
    $\kappa\simeq 0.63$ with a standard error $0.11$.  Values of other
    parameters used in the simulations are $K_{d}=r_{d}=u_{d}=1$,
    $K_{c}=0.1$. See text and~\cite{Supplement} for more details.}
\label{Fig. fig_3} 
\end{figure}

In this work, we address this question for one-dimensional driven open quantum
systems. To this end, we establish the complete phase diagram of 1D driven open
condensates (DOC) via numerical simulations in combination with analytic
analysis, and explain its basic structure on the basis of the 1D cKPZ equation
(cf.~Fig.~\ref{Fig. Phase_diagram_and_key_properties}).  It can be traced back
to the behavior of dynamical topological defects, namely space-time vortices
(Fig.~\ref{Fig. Phase_diagram_and_key_properties}(b)). More specifically, we
find: (i) The emergence of a new time scale $t_{v}$ in the long time behavior of
the temporal coherence function. At weak noise level $\sigma$, it is
exponentially large, $t_{v}\propto e^{B/\sigma}$ ($B$ a non-universal positive
constant), reflecting an exponentially suppressed but finite space-time vortex
density (Fig.~\ref{Fig. Phase_diagram_and_key_properties}(c)).  The subsequent
asymptotic regime $t\gg t_{v}$ is characterized by a disordered, exponentially
decaying first order temporal coherence function. The exponential dependence on
the inverse noise level corroborates previous results on the observability of
KPZ physics in 1D, where the crossover scale from stretched exponential
equilibrium diffusive to non-equilibrium KPZ scaling behaves algebraically as
$t_{*}\propto\sigma^{-2}$~\cite{Natermann_1d_cross_over_time_scaling} at weak
noise level, ensuring generically $t_{*}\ll t_{v}$~\cite{comment_on_2d_L_v}.
(ii) We identify a new intrinsic non-equilibrium phase of the cKPZ equation,
with the following key signatures: (a) A first order transition at low noise
level from a regime of exponentially low to high space-time vortex density
(cf.~Fig.~\ref{Fig. Phase_diagram_and_key_properties}(d)). (b) Strongly
non-thermal behavior of the momentum distribution of quasiparticles
$n_{q}\propto q^{-\gamma}$ at large momentum $q$
(cf.~Figs.~\ref{Fig. fig_3}(a) and \ref{Fig. fig_3}(b)), with $\gamma$ significantly
deviating from the value for thermal behavior ($\gamma\sim2$). Such a strong
scaling behavior is reminiscent of turbulence, which usually is a transient
phenomenon that occurs in specifically initialized systems undergoing purely
Hamiltonian dynamics without external drive~\cite{Gasenzer_2012_1D_Bose_gas}. In
contrast, it occurs in stationary state in our case.

\emph{Microscopic model}.-- We describe the dynamics in terms of the
SCGLE with complex Gaussian white noise of zero mean, as
appropriate for experiments with exciton-polariton
systems~\cite{Polarition_exp_1,Polarition_exp_2,Polarition_exp_3}. It reads in
1D~\cite{Carusotto2013} (units $\hbar=1$)
\begin{eqnarray}
\frac{\partial}{\partial t}\psi & = & \left[r+K\frac{\partial^{2}}{\partial x^{2}}+u|\psi|^{2}\right]\psi+\zeta,\label{eq:Dimension_less_SCGLE}
\end{eqnarray}
with $r=r_{d}+ir_{c}$, $K=K_{d}+iK_{c}$, $u=-u_{d}-iu_{c}$,
$\langle\zeta(x,t)\zeta(x',t')\rangle=0$,
$\langle\zeta^{*}(x,t)\zeta(x',t')\rangle=2\sigma\delta(x-x')\delta(t-t')$.
$r_{c}$ and $u_{c}$ are the chemical potential and the elastic collision
strength, respectively. $r_{d}=\gamma_{p}-\gamma_{l}$ is the difference between
the single particle loss $\gamma_{l}$ and incoherent pump $\gamma_{p}$. For the
existence of a condensate in the mean field steady state, $r_{d}>0$. $u_{d}$ is
the positive two-particle loss rate; $K_{c}=1/(2m_{\mathrm{LP}})$ with
$m_{\mathrm{LP}}$ the effective polariton mass and $K_{d}$ a diffusion
constant. We obtained most of the numerical results presented in this work by
solving Eq.~\eqref{eq:Dimension_less_SCGLE} using the same approach as in
Ref.~\cite{cri_1d_dri_con} (see also~\cite{Supplement} for
  technical details) and we set $r_{d}=K_{d}=1$; hence, $t$ and $x$ are
measured in units of $r_{d}^{-1}$ and $\sqrt{K_{d}}$. If not specified
otherwise, we used $10^{3}$ stochastic trajectories to perform ensemble
averages.

\emph{Low frequency effective description.-- } In the absence of phase defects,
the low frequency dynamics of the system is effectively described by the KPZ
equation~\cite{KPZ_equation} for the phase of the condensate field,
$\partial_{t}\theta=D\partial_{x}^{2}\theta+\frac{\lambda}{2}\left(\partial_{x}\theta\right)^{2}+\xi$~\cite{Altman_2d_driven_SF_2015}. Here,
$D$ describes phase diffusion, and $\xi$ is a Gaussian white noise of strength
$2\sigma_{\mathrm{KPZ}}$. The non-linearity $\lambda$ is a direct,
single-parameter measure for the deviation from equilibrium conditions, with
$\lambda=0$ in the presence of detailed
balance~\cite{Altman_2d_driven_SF_2015}. In order to properly account for the
compactness of the phase and to allow for a description of phase defects, we
work with a lattice regularized version of the KPZ equation (cKPZ), which can be
straightforwardly derived from a spatially discretized SCGLE on a 1D lattice
with spacing $\Delta_{x}$~\cite{Supplement}. It reads
\begin{equation}
\partial_{t}\theta_{i}=\sum_{j=i\pm1}\left[-\bar{D}\sin\left(\theta_{i}-\theta_{j}\right)+\bar{\lambda}\sin^{2}\!\!\left(\frac{\theta_{i}-\theta_{j}}{2}\right)\right]+\bar{\xi}_{i},\label{eq:NE-XY_model}
\end{equation}
with $\theta_{i}(t)\equiv\theta(i\Delta_{x},t)$, $\bar{\xi}_{i}(t)$ being
Gaussian white noise with
$\langle\bar{\xi}_{i}(t)\bar{\xi}_{i'}(t')\rangle=2\bar{\sigma}_{\mathrm{KPZ}}\delta(t-t')\delta_{ii'}$,
$\bar{D}=D/\Delta_{x}^{2}$, $\bar{\lambda}=\lambda/\Delta_{x}^{2}$,
$\bar{\sigma}_{\mathrm{KPZ}}=\sigma_{\mathrm{KPZ}}/\Delta_{x}$, where
$D=\frac{u_{c}K_{c}}{u_{d}}+K_{d},\lambda=2(\frac{u_{c}K_{d}}{u_{d}}-K_{c}),$
and $\sigma_{\mathrm{KPZ}}=\sigma\frac{u_{c}^{2}+u_{d}^{2}}{2r_{d}u_{d}}$. The
KPZ equation is reproduced upon assuming that phase fluctuations are small, and
taking the continuum limit. The crucial difference between the non-compact
continuum KPZ and the compact KPZ equation is revealed by the number of
independent scales in the problem, which originates from the compactness of the
phase. Indeed, by rescaling $t$, $\xi$, and $\theta$ in the continuum case,
there is only \emph{one} tuning parameter given by
$g\equiv\lambda(\sigma_{\mathrm{KPZ}}/2D^{3})^{1/2}$. In contrast, for the cKPZ
equation, we can rescale $t$ and $\bar{\xi}_{i}$ but \emph{not} the phase field
$\theta_{i}$ due to its compactness, resulting in \emph{two} independent tuning
parameters. Rescaling amounts to replacing
$\bar{D}\to1,\bar{\lambda}\to\tilde{\lambda}=\lambda/D,\bar{\sigma}\to\tilde{\sigma}=\Delta_{x}\sigma_{\mathrm{KPZ}}/D$
in Eq.~\eqref{eq:NE-XY_model}. The rescaled equations with $\tilde{\lambda}$ and
$-\tilde{\lambda}$ are equivalent, therefore we further redefine
$\tilde{\lambda}\equiv|\lambda/D|$.
The fact that $\tilde{\lambda}$ and $\tilde{\sigma}$ are \emph{two} independent tuning parameters 
suggests that there must exist new physics associated with
changing each of them, beyond the physics of the KPZ equation,
where changing the nonequilibrium strength $\tilde{\lambda}$ and noise strength
$\tilde{\sigma}$ are equivalent to changing the single parameter $g$. Indeed, as
we shall see in the following, the noise strength is associated with a new time
scale in the 1D DOC giving rise to an additional scaling regime in dynamical
correlation functions. Moreover, the non-equilibrium strength $\tilde{\lambda}$
can drive the system to a new non-equilibrium vortex turbulent phase via a first
order transition at low noise level.

\emph{Space-time vortex driven crossover for $\tilde{\lambda}/\tilde{\lambda}^{*}<1$.--} The prime observable that distinguishes a 1D DOC from its equilibrium counterpart~\cite{Wouters_1d_temproal,cri_1d_dri_con} is the autocorrelation function $C_{t}^{\psi}(x;t_{1},t_{2})\equiv\langle\psi^{*}(x,t_{1})\psi(x,t_{2})\rangle$. Its long time behavior is determined by fluctuations of the phase, $C_{t}^{\psi}(x;t_{1},t_{2})\propto e^{-c\Delta_{\theta}(t_{1}-t_{2})}$, where $c$ is a non-universal constant, and $\Delta_{\theta}(t_{1}-t_{2})\equiv\frac{1}{L}\int_0^L dx \langle \left[\theta(x,t_{1})-\theta(x,t_{2})\right]^{2}\rangle -\langle \theta(x,t_{1})-\theta(x,t_{2})\rangle ^{2}$ \cite{foot_note_denifition_Delta_theta}, with $L$ being the spatial size of the system. Previous numerical studies~\cite{Wouters_1d_temproal,cri_1d_dri_con} confirmed KPZ scaling $\Delta_{\theta}(t_{1}-t_{2})\propto\left|t_{1}-t_{2}\right|^{2\beta}$ with $\beta=1/3$~\cite{comment_on_beta_eq_value} \emph{in the regime of weak noise}, defined by the absence of space-time vortices (cf.~Fig.~\ref{Fig. Phase_diagram_and_key_properties}(b)) in the spatio-temporal extent of the numerical experiments. Increasing the noise level leads to proliferation of space-time vortices, which in turn strongly affect the temporal coherence of DOCs as we describe in the following.

\begin{figure}
\includegraphics[width=3.4in]{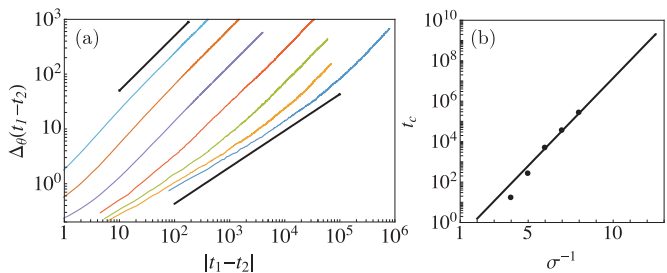}
\caption{(Color online) (a) Temporal phase fluctuations $\Delta_{\theta}(t_{1}-t_{2})$ for different noise levels on a double-logarithmic scale. The upper and lower black solid lines correspond to linear, ``disordered'' scaling $\propto\left|t_{1}-t_{2}\right|$ and KPZ scaling $\propto\left|t_{1}-t_{2}\right|^{2/3}$, respectively. From left to right, curves in between the two black solid lines are obtained for $\sigma^{-1}$ from $2$ to $8$, while the other parameters are $K_{d}=r_{d}=u_{d}=1$, $r_{c}=u_{c}=0.1$, and $K_{c}=3$. The system size $L$ is chosen large enough to make finite size effects negligible. (b) Crossover time scale $t_{c}$ as a function of inverse noise strength on a semi-logarithmic scale. The straight line is a linear fit to the data points corresponding to the lowest three noise strengths.}
\label{Fig. Phase_fluctuation_and_Crossover_time_scale} 
\end{figure}

Figure~\ref{Fig. Phase_fluctuation_and_Crossover_time_scale}(a) shows
$\Delta_{\theta}(t_{1}-t_{2})$ for moderate noise strengths. For the lowest
value $\sigma=8^{-1}$, KPZ scaling can be observed in a wide range
$5\times10^{3}\lesssim\left|t_{1}-t_{2}\right|\lesssim5\times10^{4}$, after
which $\Delta_{\theta}(t_{1}-t_{2})$ grows linearly with time. Accordingly, the
autocorrelation function $C_{t}^{\psi}(x;t_{1},t_{2})$ exhibits a crossover from
stretched-exponential to simple exponential decay at long
times~\cite{Supplement}. At weak noise, the crossover time $t_{c}$, defined as
the point where the gradient of $\Delta_{\theta}(t_{1}-t_{2})$ on a double
logarithmic scale exceeds $0.9$, increases exponentially with the inverse noise
level, i.e., $t_{c}\propto e^{C/\sigma}$, where $C$ is a positive constant. This
is shown in Fig.~\ref{Fig. Phase_fluctuation_and_Crossover_time_scale}(b). The
fast growth of phase fluctuations and the associated decoherence of DOCs for
strong noise and at long times is due to space-time vortices. Numerical evidence
for this connection is presented in
Fig.~\ref{Fig. Phase_diagram_and_key_properties}(c), where a linear fit on the
semi-logarithmic scale clearly demonstrates that the space-time vortex density
behaves as $P_v \propto e^{-A/\sigma}$ (see~\cite{Supplement} for
  details on the numerical calculation of $P_v$) in the range
$6\lesssim\sigma^{-1}\lesssim13$, while saturation sets in at even higher noise
strengths. This exponential determining $P_v$ is the same as the one appearing
in $t_{c}$ and can be interpreted as the statistical weight of space-time vortex
configurations in an equivalent 1+1D static equilibrium description of the
problem. Formally, this 1+1D static equilibrium description can be established
by rewriting the cKPZ equation as a functional integral over space-time
configurations of the phase field as
$Z=\int\mathcal{D}[\theta]e^{-\mathcal{H}[\theta]/T}$~\cite{Supplement,Golubovic},
with the Boltzmann factor $e^{-\mathcal{H}[\theta]/T}$ determining the
probability with which a particular solution $\theta(x,t)$ is encountered in a
numerical integration of Eq.~\eqref{eq:NE-XY_model} from time $t_{0}$ to $t_{1}$
in a system of size $L$. The temperature is $T=4\sigma_{\mathrm{KPZ}}$, and the
effective Hamiltonian (going back to a spatial continuum notation) is given by
$\mathcal{H}[\theta]=\int_{t_{0}}^{t_{1}}dt\int_{0}^{L}dx\left[\partial_{t}\theta-D\partial_{x}^{2}\theta-\frac{\lambda}{2}\left(\partial_{x}\theta\right)^{2}\right]^{2}$. Regarding
$z=t$ as a spatial coordinate, this is the Hamiltonian of a 2D smectic $A$
liquid crystal~\cite{Golubovic}. Crucially, one can show that the
  energy $E_{v}=\mathcal{H}[\theta_{v}]$ for a single-vortex configuration
  $\theta_{v}(x,t)$ is \emph{finite}~\cite{comment_on_finite_vx_energy},
indicating that at any finite noise strength there is a non-zero density of
noise activated vortices $P_v\propto e^{-E_{v}/T}$. Deviations
from this exponential dependence in
Fig.~\ref{Fig. Phase_diagram_and_key_properties}(c) for strong noise are due to
the interaction energy of vortices, which becomes important at higher densities.

The exponentially small but finite space-time vortex density entails a time
scale $t_{v}$ ~\cite{comment_on_phase_slip_rate_eq_case} characterizing the
average temporal distance between vortices,
$t_{v}\propto P_v^{-3/5}\propto
e^{\frac{3A}{5}\sigma^{-1}},\,\mathrm{if}\,\sigma\ll1$~\cite{foot_note_detail_explanation_tv}.
Then, using a simple phase random walk model, one can show that space-time
vortices occurring at multiples of $t_{v}$ lead to linear growth of
$\Delta_{\theta}(t_1-t_2)$, hence disordered behavior of the phase correlations,
beyond a time scale $t_{c}=\mathcal{O}(t_{v})$~\cite{Supplement}.

\emph{First order transition and vortex turbulence for
  $\tilde{\lambda}/\tilde{\lambda}^{*}>1$.--} 
We now investigate the strongly non-equilibrium regime at large
$\tilde{\lambda}$, where most of the results presented in the following are obtained from direct simulations of the SCGLE. 
Figure~\ref{Fig. Phase_diagram_and_key_properties}(d) shows
the dependence of the vortex density $P_v$ on $\tilde{\lambda}$ at weak
noise ($\sigma=10^{-2}$). At small $\tilde{\lambda}$, $P_v$ is exponentially small in line with
the discussion above. However, when tuning above a critical strength
$\tilde\lambda^{*} \simeq 3.23$, the vortex density undergoes a sudden jump by around
$10$ orders of magnitude, indicating a sharp first order transition at low noise
level~\cite{comment_on_1st_order_transition}. 

Clear signatures of the transition can also be seen in the momentum distribution
function $n_{q}\equiv\langle\psi^{*}(q)\psi(q)\rangle$, which is accessible in
experiments with exciton-polaritons. At large $q$, it behaves as
$n_{q}\propto q^{-\gamma}$
(cf.~Figs.~\ref{Fig. fig_3}(a) and~\ref{Fig. fig_3}(b)). Across the transition the
value of $\gamma$ undergoes a jump from $\gamma\simeq2$, which is characteristic
of noise activated vortices~\cite{Gasenzer_2012_1D_Bose_gas}, to
$\gamma\simeq5$ (with a weak dependence on parameters). Such strong scaling
behavior is reminiscent of turbulence and we thus refer to this phase as vortex
turbulence (VT).

The physical origin of the VT phase and the associated first order transition
can be traced back to a dynamical instability triggered by $\tilde \lambda$:
consider the dynamical equations for the phase \emph{differences} between
nearest neighboring sites, $\Delta_{i}\equiv\theta_{i}-\theta_{i+1}$, at zero
noise, which assumes the form
$\partial_{t}\Delta_{i}\simeq-(2\Delta_{i}-\Delta_{i+1}-\Delta_{i-1})+\tilde{\lambda}\left(\Delta_{i-1}^{2}-\Delta_{i+1}^{2}\right)/4$
when $\Delta_{i}$ is small. The first term, originating from the diffusion term
in the cKPZ, is a restoring force which attenuates phase differences, while the
second term amplifies them. This causes the dynamical instability for large
$\tilde{\lambda}$ and induces vortices without resorting to noise induced
excitation. The dynamical instability is thus triggered on short scales. 
Moreover, this mechanism can also occur in higher dimensions.  Taking
into account the exponential suppression of noise activated
vortices at low noise and the fact that the vortex generation due to the
dynamical instability is insensitive to weak noise, this rationalizes the
existence of a first order transition tuned by $\tilde{\lambda}$.
  Increasing the noise strength, Figs.~\ref{Fig. fig_3}(c)
  and~\ref{Fig. fig_3}(d) show that $\tilde\lambda^*$ does not depend on
  $\sigma$ within the numerical accuracy of our simulations
  ($\tilde\lambda^*=3.23$ with an error bar of $\pm0.07$). The transition line
  terminates at $\sigma^*\simeq 0.014$ at an apparent second order critical
  point. It is characterized by a vanishing jump $\Delta P_v$, whose derivative
  with respect to $\sigma$ diverges according to
  $\propto (\sigma-\sigma^*)^{-\kappa}$ with $\kappa\simeq 0.63$.

The discussion of the VT transition reveals it to be rooted in the
non-equilibrium nature of the cKPZ
equation~\cite{comment_on_deterministic_CGLE_work}. Numerically solving this equation at
low noise level, we obtain $\tilde{\lambda}^{*} \simeq 20$. This can be compared
to numerical simulations of the SCGLE at weak noise, which yield the
significantly reduced value $\tilde{\lambda}^{*} \simeq 3.23$. The discrepancy is
due to the mutual feedback between phase and density fluctuations
  present in SCGLE: the former can cause the latter via a phase-density
coupling term proportional to the diffusion constant $K_d$
(cf.~Eq.~\eqref{eq:Dimension_less_SCGLE}), which in turn facilitates strong
phase fluctuations, gives rise to vortex creation and consequently
  causes the critical value $\tilde{\lambda}^{*}$ in the SCGLE to depend on
  $K_d$.

\emph{Experimental observability.--} 
In exciton-polaritons~\cite{Polarition_exp_1d}, typically the diffusion constant
$K_d \ll K_{c}=1/2m_{LP}$ (cf.~Eq.~\eqref{eq:Dimension_less_SCGLE}). Then, in
line with previous results~\cite{cri_1d_dri_con,Wouters_1d_temproal}, KPZ
scaling should be observable due to the exponential suppression of unbound
vortices at low noise and weak non-equilibrium strength. However, 
we note that a relatively large $\tilde \lambda$ (but below the VT transition) 
is generally favorable in order to overcome the influences from potential inhomogeneities and the finite system size. 
On the other hand, $K_c \simeq 0$ in recently realized exciton-polariton condensates formed in the
flat band of a 1D Lieb lattice of micropillar optical cavities~\cite{Baboux_2016_flat_band}. 
In this case, the maximum achievable non-equilibrium strength is 
$\tilde{\lambda}_\mathrm{max} \sim 2K_{d}/K_{c}$, indicating that $\tilde\lambda$ can be 
tuned large to make the VT phase accessible. As explained above, the distinct features of this phase are revealed in
the momentum distribution and can thus be obtained by momentum resolved correlation
measurements~\cite{Carusotto2013}. Alternatively, explicit engineering of a
large diffusion constant could be achieved in 1D arrays of microwave resonators
coupled to superconducting qubits~\cite{1D_quantum_Neq_critical}.
More generally, DOC systems with relatively large $K_d/K_c$ are 
favorable to observe the VT phase, since the associated mutual feedback between phase and density fluctuations makes it easier to trigger the dynamical instability.

\emph{Conclusions}.-- The non-equilibrium phase diagram of one-dimensional
driven open condensates is crucially shaped by space-time vortices, as the
relation to the compact KPZ equation reveals: at weak non-equilibrium strength,
they govern the asymptotic behavior of the temporal correlation functions,
however only beyond an exponentially large crossover time scale. This protects
KPZ physics and suggests its observability in current exciton-polariton
experiments. Moreover, these defects cause the existence of a new phase under
strong non-equilibrium conditions, stationary vortex turbulence. We believe that
our predictions will stimulate both further theoretical research on the cKPZ
equation, especially in higher dimensions and in the context of stationary
turbulence, as well as experimental efforts in searching for these two
non-equilibrium phases. Other intriguing directions for future research are the study of nucleation dynamics and phase coexistence in the vicinity of the transition to VT, and the critical behavior of the second order end point using the techniques developed in~\cite{unbinding_XPC_2d_2}.

\emph{Note added.}-- Upon completion of this manuscript, we became aware of the work by Lauter \textit{et al.}~\cite{Marquardt_2016}, reporting a related dynamical instability in arrays of coupled phase oscillators.

\begin{acknowledgments}
  We thank I. Carusotto, B. Kim, E. Altman, D. Huse, J. Toner and S. Mathey for
  useful discussions, and the Center for Information Services and High
  Performance Computing (ZIH) at TU Dresden for allocation of computer
  time. This work was supported by German Research Foundation (DFG) through ZUK
  64, through the Institutional Strategy of the University of Cologne within the
  German Excellence Initiative (ZUK 81) and by the European Research Council via
  ERC Grant Agreement n. 647434 (DOQS). L.~S. acknowledges funding through the
  ERC synergy grant UQUAM.
\end{acknowledgments}


\begin{thebibliography}{10}
\bibitem{foot_note_denifition_Delta_theta} In the definition of $\Delta_{\theta}(t_{1}-t_{2})$, the spatial average over the system size $L$ is taken for convenience and does not affect the universal scaling behavior for large time differences $\left|t_{1}-t_{2}\right|$.
\bibitem{comment_on_phase_slip_rate_eq_case} The rate $1/t_{v}$ at which phase slips occur and hence the vortex density $P_v$ for a system in thermal equilibrium (described by Eq.~\eqref{eq:Dimension_less_SCGLE} with $r_{c}=K_{c}=u_{c}=0$) was determined in the context of narrow superconducting channels in Refs.~\cite{LAMH}. Phase slips are fluctuations which reduce the current and thus leads to a finite resistance of the channel. More formally, during a phase slip the system changes its state from one local minimum of the free energy to a neighboring one that differs in current by $2 \pi/L$. At low noise (set by the temperature $T$), the rate at which such fluctuations occur is proportional to $\exp(-\delta F/T)$ where $\delta F$ is the height of the free energy barrier separating the local minima. Out of thermal equilibrium, however, the notion of a free energy becomes void and the theory developed in Refs.~\cite{LAMH} cannot be applied directly. Then, the approach presented in the main text is needed to obtain the exponential factor.
\bibitem{foot_note_detail_explanation_tv} The mean temporal and spatial distances between vortices, denoted as $t_{v}$ and $x_{v}$ respectively, are given in terms of the vortex density by $t_{v}x_{v}=1/P_v$. Inserting here the relation $t_{v}\sim x_{v}^{z}$ with the dynamical exponent $z=3/2$, which is valid in the scaling regime of the non-compact KPZ equation, leads to $t_{v}\propto P_v^{-3/5}$.
\bibitem{comment_on_1st_order_transition} $P_v$ is finite on both sides of the transition, i.e.,
the long-time decay of the autocorrelation function is exponential on both
sides. The phases are distinct by the density of vortices---similar to the
liquid-gas transition in thermal equilibrium.
\end{thebibliography}
\end{document}


\title{Supplemental Material for ``Space-time vortex driven crossover and
vortex turbulence phase transition in one-dimensional driven open condensates''}

\author{Liang He$^{1,2}$, Lukas M. Sieberer$^{3,4}$ and Sebastian Diehl$^{1,2}$}

\affiliation{$^{1}$Institute for Theoretical Physics, Technical University Dresden,
D-01062 Dresden, Germany}

\affiliation{$^{2}$Institute for Theoretical Physics, University of Cologne,
D-50937 Cologne, Germany}

\affiliation{$^{3}$Department of Condensed Matter Physics, Weizmann Institute
of Science, Rehovot 7610001, Israel}

\affiliation{$^{4}$Department of Physics, University of California, Berkeley,
California 94720, USA}

\maketitle

\section{Derivation of the compact KPZ equation (cKPZ) and its effective 2D
static model}

Here we provide some details on the derivation of the compact KPZ
equation in one spatial dimension, and the equivalent equilibrium
description in terms of an effective Hamiltonian.

\subsection{Compact KPZ equation}

Mathematically, the compactness of the phase field can be formulated
as a ``discrete'' local gauge symmetry of the SCGLE in the amplitude-phase
representation of the field $\psi(x,t)\equiv\rho(x,t)e^{i\theta(x,t)}$.
Namely, if $\rho(x,t)e^{i\theta(x,t)}$ is a solution of the SCGLE,
then $\rho(x,t)e^{i\left[\theta(x,t)+2\pi n(x,t)\right]}$ is also
a solution, where $n(x,t)$ is an integer valued function. This symmetry
reflects the simple physical requirement that any choice of the space-time
local definition $\theta(x,t)$ by shifting $2\pi n(x,t)$ must not
change the physical results. Therefore, it is crucial to require the
effective description to respect this symmetry.

The discrete nature of the gauge transformation indicates that both
the temporal and spatial derivative of the phase field have to be
taken with care when one tries to derive the dynamical equation for
the phase field from the SCGLE. This is due to the fact that the existence
of the derivatives for any physical solution of the phase field is
not always guaranteed. To see this point, let us consider a solution
of the SCGLE, $\psi(x,t)=\rho(x,t)e^{i\theta(x,t)}$ with $\theta(x,t)$
having well defined $\partial_{t}\theta$ and $\partial_{x}\theta$
at any space-time point. For a generic gauge transformed solution
$\tilde{\psi}=\rho(x,t)e^{i\tilde{\theta}(x,t)}$ with $\tilde{\theta}(x,t)=\theta(x,t)+2\pi n(x,t)$,
one immediately sees that $\partial_{t}\tilde{\theta}$ or $\partial_{x}\tilde{\theta}$
is not well defined if around a certain space-time point, say $(x',t')$,
$n(x',t'+0^{+})\neq n(x',t'+0^{-})$ or $n(x'+0^{+},t')\neq n(x'+0^{-},t')$.
Nevertheless, due to fact that the exponential function $e^{i\alpha}$
is a smooth function with respect to $\alpha$ for any physical solution
of the phase field, one can always choose a gauge $n(x,t)$ in such
a way that \emph{either} the temporal \emph{or} the spatial derivative
of the transformed phase field exists everywhere in the space-time
domain under consideration, but \emph{not} both.

In the following, we choose the gauge in which the temporal derivative
of the phase field exists everywhere. We choose the value of $n(x,t+0^{+})$
according to $\theta(x,t)$ in such a way that $\theta(x,t+0^{+})-\theta(x,t)$
is infinitesimal. Then, the spatial derivative of the phase field
is not well-defined everywhere. At the price of introducing a short
distance scale, we can circumvent this issue by first rewriting the
SCGLE on a discrete 1D lattice with lattice spacing $\Delta_{x}$,
such that the second order spatial derivative is replaced by the a
second order finite difference. The discretized SCGLE reads 
\begin{equation}
\partial_{t}\psi_{i}=(r+u\left|\psi_{i}\right|^{2})\psi_{i}+\frac{K}{\Delta_{x}^{2}}\left(\psi_{i+1}+\psi_{i-1}-2\psi_{i}\right)+\zeta_{i},\label{eq:Discrete_SCGLE}
\end{equation}
with $\psi_{i}(t)\equiv\psi(i\Delta_{x},t)$ and $\langle\zeta_{i}(t)\zeta_{i'}(t')\rangle=2\sigma\delta(t-t')\delta_{ii'}/\Delta_{x}$.

The rest of the derivation goes along the lines of the similar derivation
presented in~\cite{Altman_2d_driven_SF_2015}. Plugging the amplitude-phase
decomposition of $\psi$, i.e. $\psi_{i}(t)=(M_{0}+\chi_{i}(t))e^{i\theta_{i}(t)}$
with $M_{0}$ the homogeneous mean-field solution, into~(\ref{eq:Discrete_SCGLE}),
we arrive at a coupled discretized equation of motion (EOM) of amplitude
and phase fluctuations,\begin{widetext} 
\begin{align}
\partial_{t}\chi_{i} & =(r_{d}-u_{d}\left[M_{0}+\chi_{i}\right]^{2})\left[M_{0}+\chi_{i}\right]\\
 & +\frac{K_{d}}{\Delta_{x}^{2}}\left(\left[M_{0}+\chi_{i+1}\right]\cos\left(\theta_{i+1}-\theta_{i}\right)+\left[M_{0}+\chi_{i-1}\right]\cos\left(\theta_{i-1}-\theta_{i}\right)-2\left[M_{0}+\chi_{i}\right]\right)\nonumber \\
 & -\frac{K_{c}}{\Delta_{x}^{2}}\left(\left[M_{0}+\chi_{i+1}\right]\sin\left(\theta_{i+1}-\theta_{i}\right)+\left[M_{0}+\chi_{i-1}\right]\sin\left(\theta_{i-1}-\theta_{i}\right)\right)+\Re[\zeta_{i}e^{-i\theta_{i}}],\nonumber \\
\partial_{t}\theta_{i} & =(r_{c}-u_{c}\left[M_{0}+\chi_{i}\right]^{2})\\
 & +\frac{K_{c}}{\Delta_{x}^{2}}\left(\frac{\left[M_{0}+\chi_{i+1}\right]}{\left[M_{0}+\chi_{i}\right]}\cos\left(\theta_{i+1}-\theta_{i}\right)+\frac{\left[M_{0}+\chi_{i-1}\right]}{\left[M_{0}+\chi_{i}\right]}\cos\left(\theta_{i-1}-\theta_{i}\right)-2\right)\nonumber \\
 & +\frac{K_{d}}{\Delta_{x}^{2}}\left(\frac{\left[M_{0}+\chi_{i+1}\right]}{\left[M_{0}+\chi_{i}\right]}\sin\left(\theta_{i+1}-\theta_{i}\right)+\frac{\left[M_{0}+\chi_{i-1}\right]}{\left[M_{0}+\chi_{i}\right]}\sin\left(\theta_{i-1}-\theta_{i}\right)\right)+\frac{\Im[\zeta_{i}e^{-i\theta_{i}}]}{M_{0}+\chi_{i}}.\nonumber 
\end{align}
We linearize in $\chi$ and eliminate it adiabatically, made possible
since its evolution is fast compared to the gapless phase degree of
freedom. From the EOM of $\chi_{i}$ we get 
\begin{align}
\chi_{i}= & -\frac{1}{2u_{d}M_{0}^{2}}\left\{ -\frac{K_{d}}{\Delta_{x}^{2}}\left(\left[M_{0}+\chi_{i+1}\right]\cos\left(\theta_{i+1}-\theta_{i}\right)+\left[M_{0}+\chi_{i-1}\right]\cos\left(\theta_{i-1}-\theta_{i}\right)-2\left[M_{0}+\chi_{i}\right]\right)\right.\nonumber \\
 & \left.+\frac{K_{c}}{\Delta_{x}^{2}}\left(\left[M_{0}+\chi_{i+1}\right]\sin\left(\theta_{i+1}-\theta_{i}\right)+\left[M_{0}+\chi_{i-1}\right]\sin\left(\theta_{i-1}-\theta_{i}\right)\right)-\Re[\zeta_{i}e^{-i\theta_{i}}]\right\} .
\end{align}
\end{widetext}
Plugging the above equation into the EOM of $\theta_{i}$, neglecting
sub-leading and non-linear amplitude fluctuations $\partial_{t}\chi_{i},\chi_{i}^{2},\chi_{i}^{3}$
also here, and keeping only the additive part of the noise, we get
the compact KPZ equation 
\begin{equation}
\partial_{t}\theta_{i}=\sum_{j=i\pm1}\left[-\bar{D}\sin\left(\theta_{i}-\theta_{j}\right)+\bar{\lambda}
  \sin^2 \! \left(\frac{\theta_{i}-\theta_{j}}{2}\right)
  \right]+\bar{\xi}_{i},\label{eq:NE-XY_model}
\end{equation}
with $ $$\theta_{i}(t)\equiv\theta(i\Delta_{x},t)$, $\bar{\xi}_{i}(t)$
being Gaussian white noise with $\langle\bar{\xi}_{i}(t)\bar{\xi}_{i'}(t')\rangle=2\bar{\sigma}_{\mathrm{KPZ}}\delta(t-t')\delta_{ii'}$,
$\bar{D}=D/\Delta_{x}^{2}$, $\bar{\lambda}=\lambda/\Delta_{x}^{2}$,
$\bar{\sigma}_{\mathrm{KPZ}}=\sigma_{\mathrm{KPZ}}/\Delta_{x}$, where
\begin{equation}
D=\frac{u_{c}K_{c}}{u_{d}}+K_{d},\quad\lambda=2\left(\frac{u_{c}K_{d}}{u_{d}}-K_{c}\right),\quad\sigma_{\mathrm{KPZ}}=\sigma\frac{u_{c}^{2}+u_{d}^{2}}{2r_{d}u_{d}}.\label{eq:KPZ_parameter}
\end{equation}

\subsection{2D static phase model $\mathcal{H}[\theta]$}

Here, we construct the effective 2D static equilibrium model for the
compact KPZ equation. Conceptually, the idea is to interpret the compact
KPZ equation as a mapping from the set of stochastic variables $\bar{\xi}(t)=\{\bar{\xi}_{i}(t)\}$
with Gaussian path probability distribution 
\begin{equation}
\mathcal{P}[\bar{\xi}]\propto e^{-\frac{1}{4\bar{\sigma}_{\mathrm{KPZ}}}\sum_{i}\int_{t_{0}}^{t_{1}}dt\,\bar{\xi}_{i}(t)^{2}}\label{eq:1}
\end{equation}
to new stochastic variables $\theta(t)=\{\theta_{i}(t)\}$. The distribution
of the new variables can be obtained using the usual relation~\cite{Graham1973}
\begin{equation}
\mathcal{P}[\bar{\xi}]\mathcal{D}[\bar{\xi}]=\mathcal{P}_{\theta}[\theta]\mathcal{D}[\theta],\label{eq:2}
\end{equation}
where $\mathcal{D}[\theta]=\prod_{i}\mathcal{D}[\theta_{i}]$ is the
functional measure. This yields 
\begin{equation}
\mathcal{P}_{\theta}[\theta]=\mathcal{P}[\bar{\xi}]\left|\frac{\mathcal{D}[\bar{\xi}]}{\mathcal{D}[\theta]}\right|\propto e^{-\mathcal{H}[\theta]/T},\label{eq:3}
\end{equation}
where $T \equiv 4 \bar{\sigma}_{\mathrm{KPZ}}$ is the effective temperature and
$\mathcal{H}[\theta]$ assumes the form
\begin{multline}
  \mathcal{H}[\theta]=\int_{t_{0}}^{t_{1}}dt\Delta_{x}\sum_{i}\left[\partial_{t}\theta_{i} \vphantom{\sin^{2}
      \! \left(\frac{\theta_{i}-\theta_{j}}{2}\right)}
  \right. \\ \left. +\sum_{j=i\pm1}\left(\bar{D}\sin(\theta_{i}-\theta_{j})-\bar{\lambda}\sin^{2}
      \! \left(\frac{\theta_{i}-\theta_{j}}{2}\right)\right)\right]^{2}.\label{eq:H2D}
\end{multline}
In deriving $\mathcal{P}_{\theta}$, we used that the Jacobian determinant
associated with the change of variables from $\bar{\xi}$ to $\theta$ is equal to
unity for a retarded regularization of the time derivative in the compact KPZ
equation. In the continuum limit, $\mathcal{H}[\theta]$ reduces to the form
shown in the main text. Finally, calculating the normalization of
$\mathcal{P}_{\theta}$ corresponds to calculating the partition function $Z$ of
the model defined by Eq.~\eqref{eq:H2D} at temperature $T$,
\begin{equation}
Z=\int\prod_{i}\mathcal{D}[\theta_{i}]\, e^{-\mathcal{H}[\theta]/T}.\label{eq:5}
\end{equation}

\section{Crossover behavior in the temporal correlation function}

A crossover behavior similar to the one seen in the temporal phase
fluctuation $\Delta_{\theta}(t_{1}-t_{2})$ is also observed in the
modulus of the condensate temporal auto correlation function

\begin{eqnarray}
C_{t}^{\psi}(x;t_{1},t_{2}) & \equiv & \langle\psi^{*}(x,t_{1})\psi(x,t_{2})\rangle.
\end{eqnarray}
In practice, by assuming spatial translational invariance of the correlation
function, we calculate the spatially averaged correlation functions,
i.e., $\bar{C}_{t}(t_{1},t_{2})\equiv L^{-1}\int_{0}^{L}dx\, C_{t}^{\psi}(x;t_{1},t_{2})$,
which is equivalent to the corresponding correlation function above
but helps to reduce the statistical error.

\begin{figure}
\includegraphics[width=2.7in]{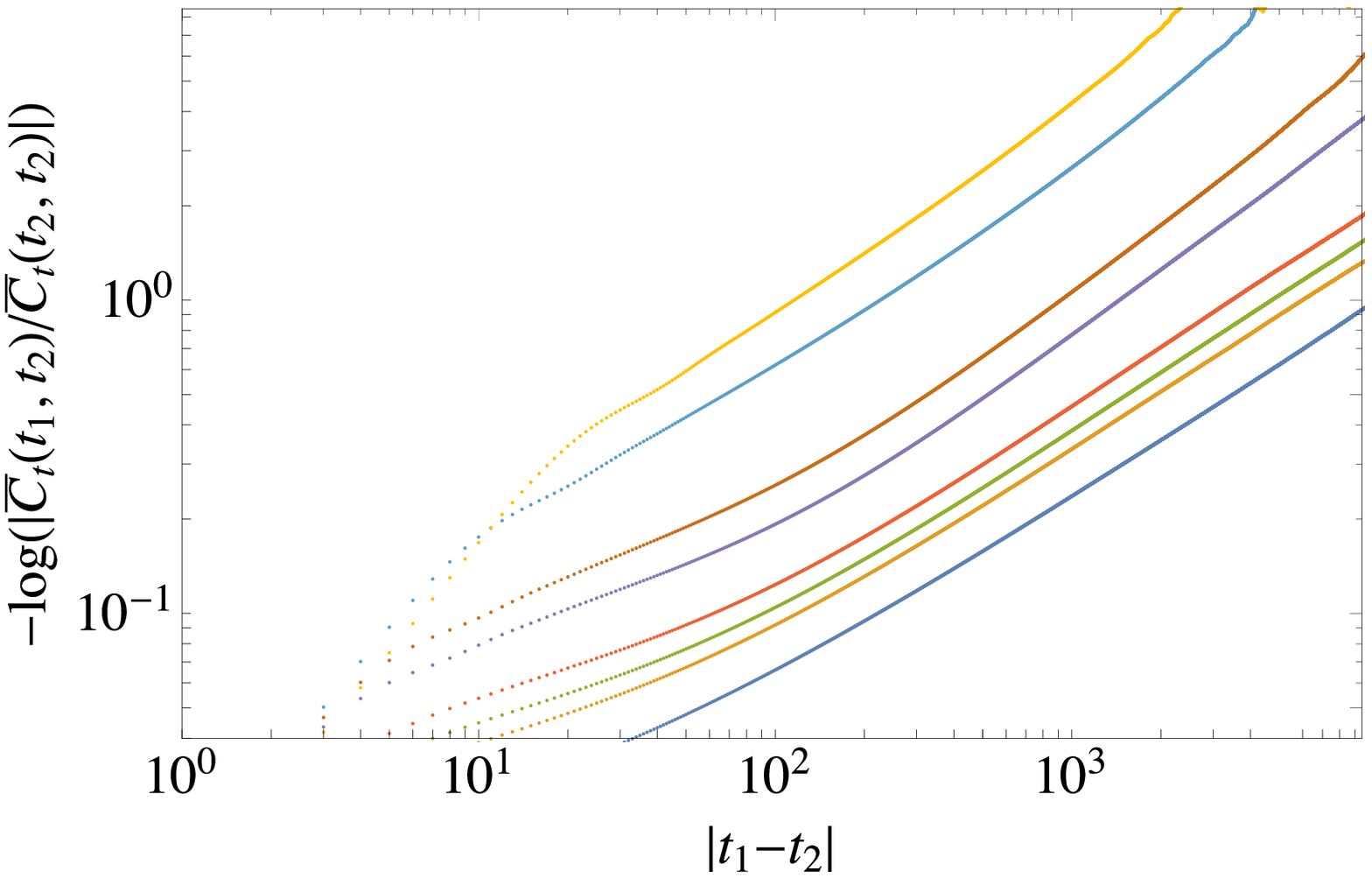}

\caption{(Color online) The dependence of $-\log\left(\left|\bar{C}_{t}(t_{1},t_{2})\right|/\left|\bar{C}_{t}(t_{2},t_{2})\right|\right)$
on $|t_{1}-t_{2}|$ for 8 different sets of parameters in the SCGLE
at linear system size $L=2^{10}$. $N_{\mathrm{Traj}}=10^{3}$ stochastic
trajectories are used to perform the ensemble average. From right
to left, the different curves corresponds to $\sigma=20^{-1},12^{-1},10^{-1},8^{-1},5^{-1},4^{-1},2^{-1},1$.
Other parameter are $K_{d}=r_{d}=u_{d}=1$, $-r_{c}=u_{c}=0.1$, and
$K_{c}=3$.}
\label{Fig. psi_correlation} 
\end{figure}

In Fig.~\ref{Fig. psi_correlation}, at linear system size $L=2^{10}$, the
dependence of
$-\log\left(\left|\bar{C}_{t}(t_{1},t_{2})\right|/\left|\bar{C}_{t}(t_{2},t_{2})\right|\right)$
on $|t_{1}-t_{2}|$ for 8 different sets of parameters in the SCGLE are shown on
a double-logarithmic scale. The exponent $\beta$ for different parameter choices
is extracted from the slope of a selected portion of the corresponding curve of
$-\log\left(\left|\bar{C}_{t}(t_{1},t_{2})\right|/\left|\bar{C}_{t}(t_{2},t_{2})\right|\right)$
on the double-logarithmic scale (cf. Fig.~\ref{Fig. psi_correlation}).  For the
noise levels $\sigma=20^{-1},12^{-1},10^{-1},8^{-1},5^{-1},$ the corresponding
slopes are extracted by performing linear fits to the data points with
$|t_{1}-t_{2}|\in[10^{3},10^{4}]$. This gives rise to $2\beta=0.61$, for all the
first four curves from below with $\sigma=20^{-1},12^{-1},10^{-1},8^{-1}$,
respectively, where we notice the typical KPZ scaling behavior, but no crossover
effect to the anticipated exponential decay. This due to the fact that the
crossover time scales corresponding to these noise levels are much larger than
the time range available in our simulations.

For the fifth curve from below with $\sigma=5^{-1}$, one notices
a considerably increased slope at the right end of the curve, clearly
indicating the crossover effect which sets in within $|t_{1}-t_{2}|\in[10^{3},10^{4}]$.
This is reflected in the extracted exponent with $2\beta=0.74$.
The anticipated exponential decay is expected to manifest itself at
even larger $|t_{1}-t_{2}|$, which is however not accessible for
our current computation resources.

For the first three curves from above with $\sigma=1,2^{-1},4^{-1}$,
since the local gradient of the curve changing noticeably up to the
last accessible data point with the largest value of $|t_{1}-t_{2}|$,
denoted as $\Delta t_{\mathrm{max}}$, the corresponding slopes are
extracted by performing linear fits to the data points of a short
segment of the curve with $|t_{1}-t_{2}|\in[\Delta t_{\mathrm{max}}e^{-1/4},\Delta t_{\mathrm{max}}]$
for $\sigma=4^{-1},2^{-1}$, and an even shorter segment of the curve
with $|t_{1}-t_{2}|\in[\Delta t_{\mathrm{max}}e^{-1/5},\Delta t_{\mathrm{max}}]$,
for $\sigma=1$. This gives rise to $2\beta=1.0$ for all the three
curves, which clearly indicates the anticipated exponential decay.
However, to further numerically verify this exponential decay in a
larger time range is beyond our currently accessible computation resources.

\section{Phase random walk model}
To see how the finite density of space-time vortices causes the exponential decay of the autocorrelation function, one can consider a simple ``phase random walk'' model in which the time evolution of the phase is assumed to be only determined by the phase jumps due to $N$ uniformly distributed vortex cores at times $nt_{v}$, where $n=1,\dotsc,N,$ with random charges $W_{n}=\pm1$ occurring with equal probability. The model can be formulated as
\begin{eqnarray}
\theta(x,t_{1}+Nt_{v})-\theta(x,t_{1})=\delta\Theta\sum_{n=1}^{N}W_{n},
\end{eqnarray}
where $\delta\Theta$ corresponds to the average amplitude of jumps of the phase field if a vortex core is crossed along the temporal direction. A straightforward calculation shows that $\Delta_{\theta}(t_{1}-t_{2})=\left(\delta\Theta\right)^{2}\left|t_{1}-t_{2}\right|/t_{v}$, i.e., space-time vortices indeed lead to disordered behavior of the phase correlations beyond a time scale $\mathcal{O}(t_{v})$

\section{Technical details in numerical simulations}

In numerical simulations, we discretize the SCGLE in space and time by
$\Delta_{x}$ and $\Delta_{t}$, respectively, and solve the discretized dynamical
equation by using the semi-implicit algorithm developed
in~\cite{Werner_1997}. Each Gaussian white noise sequence $\xi(x,t)$, which is
used in the simulation for each corresponding stochastic trajectory of
$\psi(x,t)$, is generated by using the pseudorandom number generator employing
the Mersenne Twister algorithm~\cite{MT_algorithm}. Typically, we choose
$\Delta_{x}=1$, $\Delta_{t}\in[0.01,0.1]$ in dimensionless units and the spatial
size of the system $L=2^{10}$, which is large enough to avoid finite size
effects in most simulations presented in the main text. However, for results
corresponding to the low noise levels presented in the Fig. 2 in the main text,
we have performed simulations on systems with spatial sizes up to $2^{14}$ to
avoid finite size effects that set in at long time. Typically, we simulate the
SCGLE for a time period $T\in[10^5,10^7]$ in dimensionless units, whose specific
value depends on whether physical quantities of interest have reached their
steady states. With the direct access to the time evolution of $\psi(x,t)$ field
for each stochastic trajectory simulated, the condensate temporal auto
correlation function
$C_{t}^{\psi}(x;t_{1},t_{2})\equiv\langle\psi^{*}(x,t_{1})\psi(x,t_{2})\rangle$
and momentum distribution $n_{q}\equiv\langle\psi^{*}(q)\psi(q)\rangle$ can be
calculated straightforwardly.

In order to calculate the temporal phase fluctuation function $\Delta_{\theta}(t_{1}-t_{2})\equiv\frac{1}{L}\int_{x}\left\langle \left[\theta(x,t_{1})-\theta(x,t_{2})\right]^{2}\right\rangle -\left\langle \theta(x,t_{1})-\theta(x,t_{2})\right\rangle ^{2}$,
we need to further extract the dynamics of the phase field $\theta(x,t)$. 
At each time point $t$, we first assign $\theta(x,t)$
to be the complex argument field $\arg\psi(x,t)$ and then add a multiple of $2\pi$ to the
value of $\theta(x,t)$ at each spatial point $x$ in such a way
that the phase difference between the new value of the phase at the current time point
and the value of the phase at previous time step, i.e., $\theta(x,t)-\theta(x,t-\Delta_t)$,
assumes its value within the interval $[-\pi,\pi)$. This procedure corresponds to
a choice of the gauge for the phase field that eliminates the artificial
discontinuous behavior associated with directly assigning $\theta(x,t)$
to be $\arg\psi(x,t)$. Afterwards, the temporal phase fluctuation function
$\Delta_{\theta}(t_{1}-t_{2})$ can be directly calculated. 

\begin{figure}
\includegraphics[height=1.7in]{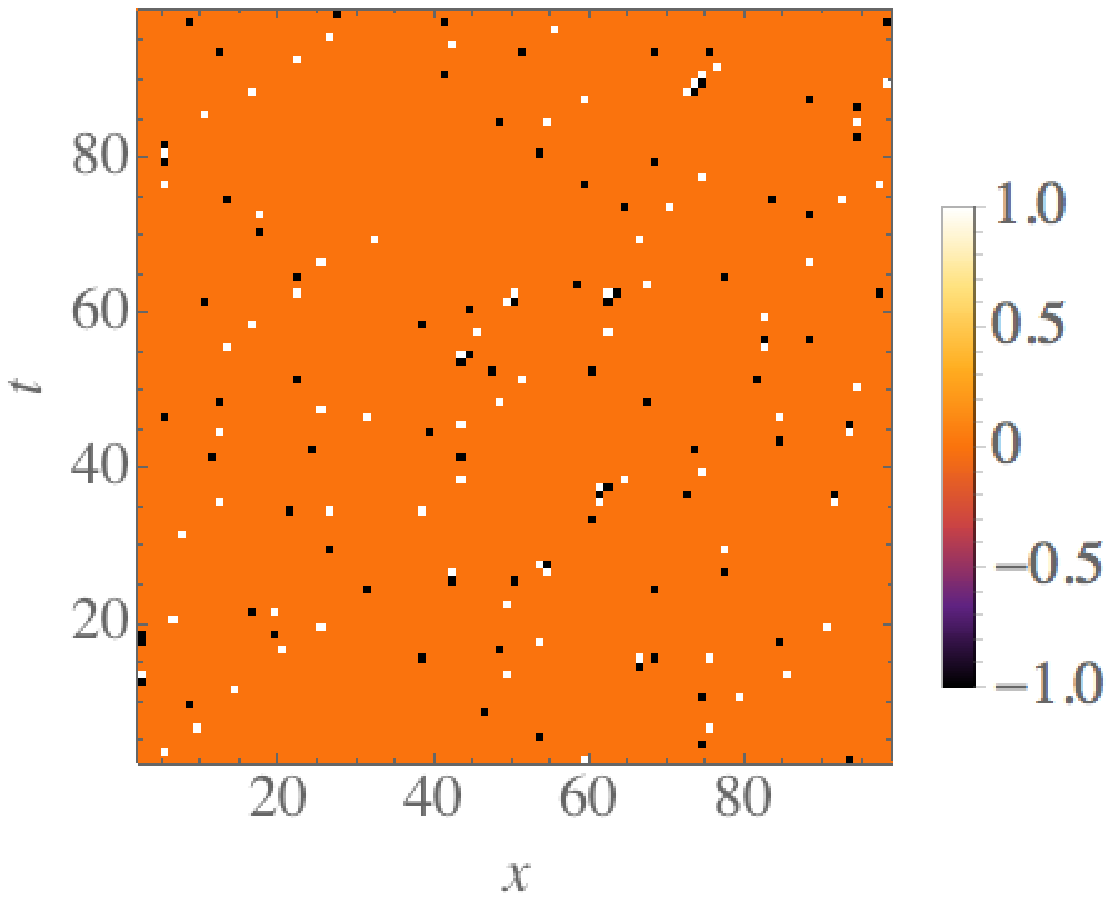}

\caption{(Color online) A snapshot of space-time vortex charge
distribution for vortex turbulence phase at large $\tilde{\lambda}$
($\tilde{\lambda}\approx4.5$) with $\tilde{\lambda}>\tilde{\lambda}^{*}$
($\tilde{\lambda}^{*}\approx3.2$) in the weak noise regime ($\sigma=0.01$
in this case). Other parameters used in the simulation are $K_{d}=r_{d}=u_{d}=1,\,r_{c}=u_{c}=3,\, K_{c}=0.1$.
Color bar stands for the charge of vortices.}

\label{Fig. vx_distri} 
\end{figure}

To calculate the vortex density $P_{v}$, for each stochastic trajectory,
we first construct the discrete phase distribution $\theta_{ij}\equiv(i\Delta x,j\Delta t)$,
choosing $\Delta x=1$ and $\Delta t=1$, on the space-time plane
of the size $L\times\Delta T$ (with $\Delta T$ usually chosen to be
$10^{3}$). and then calculate the winding number of the $\theta_{ij}$ field
on each plaquette, where a vortex is identified whenever the winding number
of the corresponding plaquette, i.e., the charge of the corresponding
vortex, is non-zero. The vortex density $P_{v}$ is obtained by calculating
$\langle N_{v}/(L\times\Delta T)\rangle$, with $N_{v}$ being the
total number of vortices on the space-time plane of the size $L\times\Delta T$
for each stochastic trajectory. In Fig.~\ref{Fig. vx_distri}, a typical snapshot of a vortex charge distribution on the space-time plane of the system in the vortex turbulence phase is shown, from which we notice the vortex charge distribution assumes an irregular structure.

\section{The first order transition line}

\begin{figure*}
\includegraphics[width=3in]{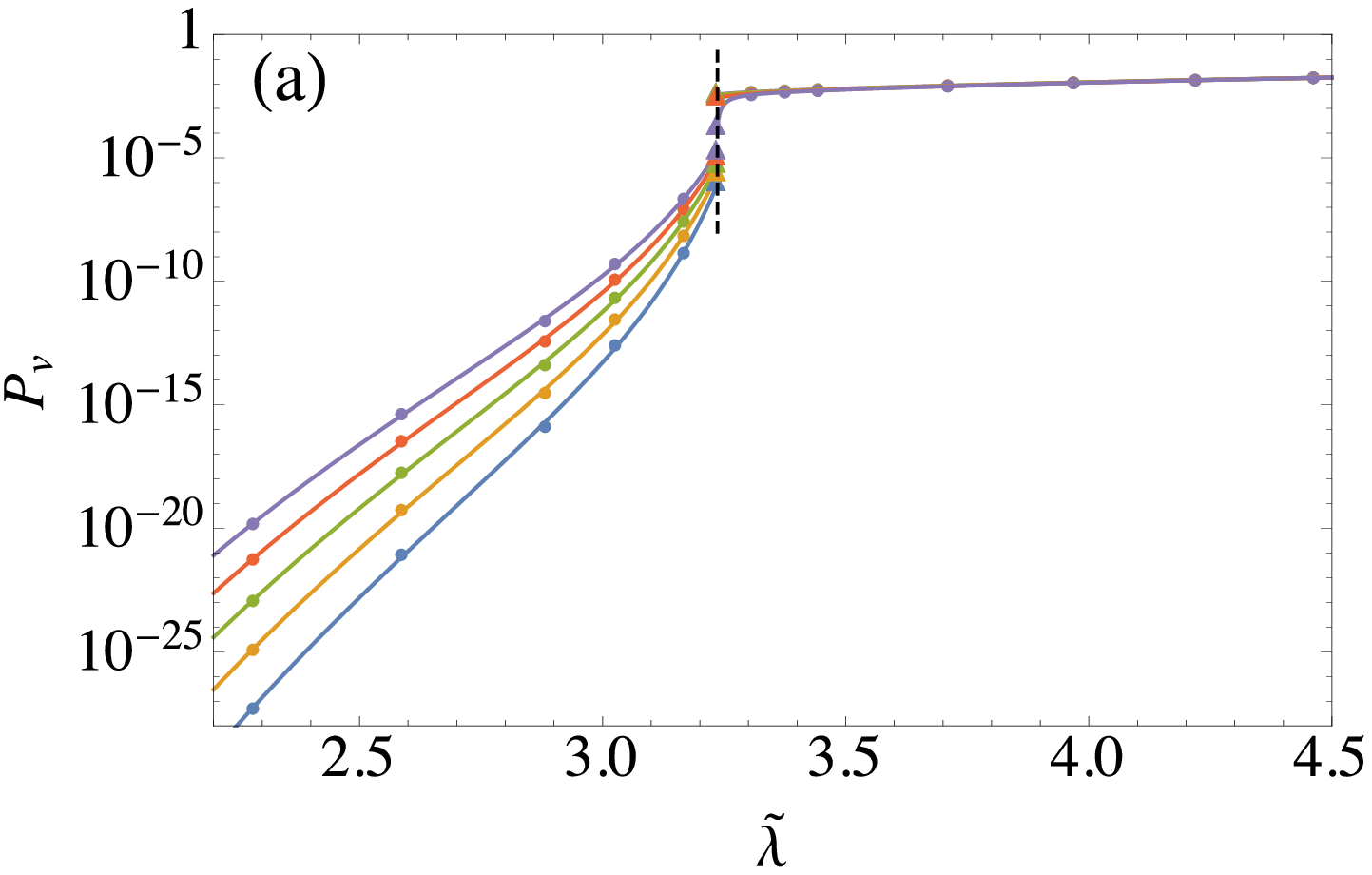}\includegraphics[width=3in]{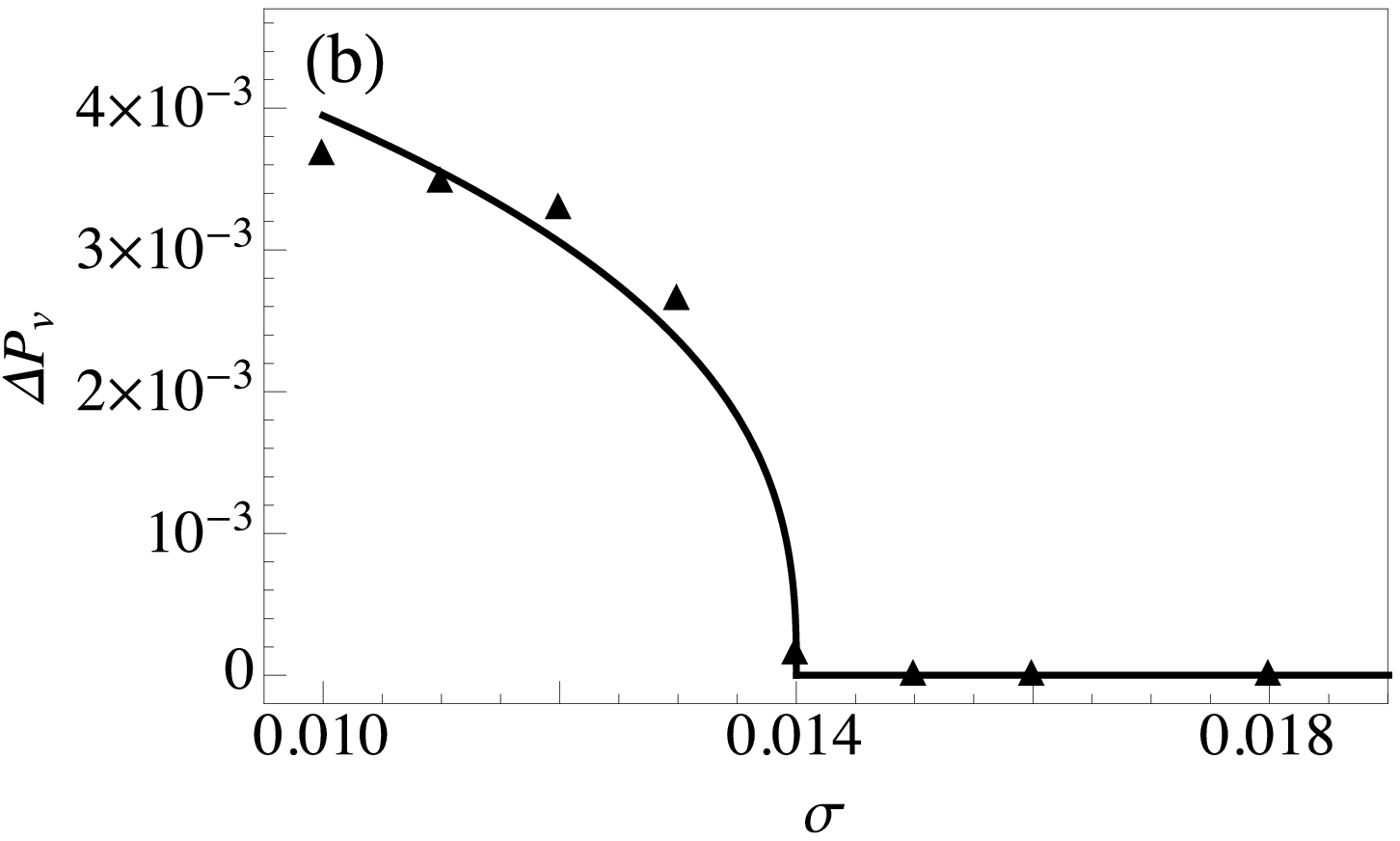}

\caption{(Color online) (a) Dependence of $P_{v}$ on the nonequilibrium strength
$\tilde{\lambda}$ at different noise level in the weak noise regime
when $\tilde{\lambda}$ is tuned across the critical value $\tilde{\lambda}^{*}$
of the first order transition. The black dashed vertical line corresponds
to the estimated value of $\tilde{\lambda}^{*}$. From down to up
(up to down), curves on the left (right) hand side of the black dashed
line correspond to noise levels $\sigma=0.01,\,0.011,\,0.012,\,0.013,\,0.014$,
respectively. The filled circles are data points obtained by numerical
simulations, while the pairs of filled triangles in the same color
at lower and upper locations correspond to the estimated values of
the left and the right limit of $P_{v}$ at $\tilde{\lambda}^{*}$,
i.e., $P_{v}(\tilde{\lambda}\rightarrow\tilde{\lambda}^{*-})$ and
$P_{v}(\tilde{\lambda}\rightarrow\tilde{\lambda}^{*+})$, respectively.
Values of other parameters used in the simulations are $K_{d}=r_{d}=u_{d}=1$,
$K_{c}=0.1$. $\tilde{\lambda}$ is tuned by changing $r_{c}=u_{c}$
from $1.4$ to $3.0$. (b) Space-time vortex density jump $\Delta P_{v}$
at the first order transition at different noise levels $\sigma$ (essentially the same plot as the one in Fig. 3(d) in the main text).
The left part of the black curve before the critical point ($\sigma\leq \sigma^*$) is a power law ($\Delta P_{v}\propto (\sigma-\sigma^*)^{1-\kappa}$) fit to the five data points on the left, which gives rise to $\kappa\simeq 0.63$ with a standard error
$0.11$. 
}
\label{Fig. P_v_vs_lambda_at_different_sigma} 
\end{figure*}

In this section we present technical details concerning the investigation of the properties
of the first order transition line. The dependence of $P_{v}$ on the
nonequilibrium strength $\tilde{\lambda}$ at different noise levels
with $\sigma=0.01,\,0.011,\,0.012,\,0.013,\,0.014,$ in the weak noise
regime are shown in Fig.~\ref{Fig. P_v_vs_lambda_at_different_sigma}(a).
At each noise level, we notice a sudden increase of $P_{v}$ by a
few orders of magnitude when $\tilde{\lambda}$ is tuned across around
$3.2$, signifying a first order transition behavior. We estimate
the critical value $\tilde{\lambda}^{*}$ for the transition at each
noise level (cf. black dashed vertical line in Fig.~\ref{Fig. P_v_vs_lambda_at_different_sigma}(a))
by the arithmetic average of the values of $\tilde{\lambda}$ for
the two data points right before and after the transition. This estimation
gives rise to $\tilde{\lambda}^{*}=3.23$ with an error bar of $\pm0.07$
for each noise level shown in Fig.~\ref{Fig. P_v_vs_lambda_at_different_sigma}(a),
indicating that, within the numerical accuracy of our simulations, $\tilde{\lambda}^{*}$
is insensitive with respect to the noise level in the weak noise regime
as shown in Fig.~3(c) in the main text.

To estimate the space-time vortex density jump at the first order
transition at each noise level, denoted as $\Delta P_{v}\equiv P_{v}(\tilde{\lambda}\rightarrow\tilde{\lambda}^{*+})-P_{v}(\tilde{\lambda}\rightarrow\tilde{\lambda}^{*-})$
with $P_{v}(\tilde{\lambda}\rightarrow\tilde{\lambda}^{*-})$ ($P_{v}(\tilde{\lambda}\rightarrow\tilde{\lambda}^{*+})$)
being the left (right) limit of $P_{v}$ at $\tilde{\lambda}^{*}$,
we first perform a rational function $R(x)\equiv(a_{0}+a_{1}x+a_{2}x^{2})/(1+b_{1}x+b_{2}x^{2})$
fit to the data points lying on the left and the right hand side of the
transition, respectively, which are shown as solid curves in Fig.~\ref{Fig. P_v_vs_lambda_at_different_sigma}(a).
Then, the value of $P_{v}(\tilde{\lambda}\rightarrow\tilde{\lambda}^{*-})$
($P_{v}(\tilde{\lambda}\rightarrow\tilde{\lambda}^{*+})$) is obtained
by performing extrapolation at $\tilde{\lambda}^{*}$ of the corresponding
curve on the left (right) hand side of the transition. From the dependence
of $\Delta P_{v}$ on the noise level $\sigma$ shown in Fig.~\ref{Fig. P_v_vs_lambda_at_different_sigma}(b) (essentially the same plot as the one in Fig. 3(d) in the main text), we notice that $\Delta P_{v}$ decreases with respect to $\sigma$
and vanishes at $\sigma^{*}\approx0.014$ with a diverging derivative of $\Delta P_{v}$ with respect to $\sigma$. 
In order to quantitatively estimate the exponent associated with the divergence of the derivative, we further perform a power law fit ($\Delta P_{v}\propto (\sigma-\sigma^*)^{1-\kappa}$) to the five data points on the left in Fig.~\ref{Fig. P_v_vs_lambda_at_different_sigma}(b), from which we extract $\kappa\simeq 0.63$ with a standard error $0.11$ and observe that $\partial \Delta P_{v} / \partial \sigma$ diverges according to $\propto (\sigma-\sigma^*)^{-\kappa}$ with $\kappa\simeq 0.63$ at $\sigma=\sigma^*$. 
These observations indicate that the first order trasition line on the $\tilde{\lambda}-\sigma$ plane
terminates at higher noise level at a second order critical point
$(\tilde{\lambda}^{*},\sigma^{*})$ as shown in Fig.~3(c) in the main text.

